\documentclass[reprint,amsmath,amssymb,aip,jcp,author-numerical]{revtex4-1}
\usepackage{graphics}
\usepackage{graphicx}
\usepackage{dcolumn}% Align table columns on decimal point
\usepackage{bm}
\usepackage{natbib}
\usepackage{hyperref}
\usepackage{color}
\begin{document}

\title{Dielectric response in the vicinity of an ion: A nonlocal and nonlinear model of the dielectric properties of water}
\author{H. Berthoumieux}

\affiliation{CNRS, UMR 7600, LPTMC, F-75005, Paris, France}
\affiliation{Sorbonne Universit\'es, UPMC Univ Paris 06, UMR 7600, LPTMC, F-75005, Paris, France}
\author{F. Paillusson }
\affiliation{School of Mathematics and Physics, University of Lincoln, LN6 7TS, UK. }
\begin{abstract}
The goal of this work is to propose a simple continuous model that captures the dielectric properties of water at the nanometric scale. We write an electrostatic energy as a functional of the polarisation field containing a term in $P^4$ and non-local Gaussian terms. Such an hamiltonian can reproduce two key properties of water: the saturation of the polarisation response of water in the presence of a strong electrostatic field and the nanometric dipolar correlations of the solvent molecules modifying the long range  van der waals interaction. This model explores thus two fundamental aspects that have to be included in implicit models of electrolytes for a relevant description of electrostatic interactions at nanometric scales. 
\end{abstract}
\maketitle

\section{Introduction} 
The dielectric properties of a medium represent a crucial model ingredient of any theory of condensed matter phase in that they underly many phenomena characteristic of the said medium: from the solvation energy \cite{Kornyshev1985,Bashford2000} and the role of ions in liquid phases \cite{Le11} to the dispersion forces between mesoscopic media of different kinds \cite{Parsegian2006}. Changing the dielectric model can have dramatic effects on the qualitative behaviour of a system indeed \cite{kornyshevsat1997, hildebrandt2004, Podgornik04,levy2012,li2014}.\newline For most liquid phases, one can assume that the dielectric response is local --- {\it i.e.} the medium dielectric polarisation field at any given point only depends on the electric field at that same point --- and linear --- {\it i.e.} the medium dielectric polarisation field and the electric field can be related by a simple electric-field-independent linear kernel ---. This is a very good approximation for systems much larger than the solvent dipolar correlation lengths and for weak electric fields but which becomes {\it a priori} insufficient for other scenarios. For example, Lab-On-Chip experiments involving micro- and nano-fluidic channels are on the rise \cite{Samiei16, siria2017}, many of the biological organelles are of nanometric size \cite{cogliati2016}, the dielectric response of the solvant in ions hydration shell can not be modeled with linear models that overestimate it \cite{levy2012,berthoumieux2018} and numbers of recent experimental and numerical measurements have confirmed that the dielectric properties of the fluid close to interfaces drastically differ from bulk properties \cite{shaff2015,fumagalli2018}. 

To make things worse, water, the most ubiquitous liquid on Earth, has been shown to display features characteristic of non-local dielectric models with bulk dipolar correlations that extend over few nanometers and oscillate. This is because water is an associated liquid structured by a network of intermolecular hydrogen bonds. The few $k_BT$ strength of the H-bond leads to correlations between the water dipoles orientations on a few hydration shells. \newline It appears all the more relevant that water does rarely comes in 'pure' form. The properties of water molecules around ions remain a topic of active research as it plays a role both at the molecular and at the nanometric level. For example, the electrostatic interaction between an ion and the water molecules of its first solvation layer is the dominant contribution to the NMR relaxation measurements of quadrupolar ions in liquid phase \cite{carof2015}.\newline It thus appears that in addition to the existing good understanding of the local and linear dielectric responses of liquids, it becomes increasingly important to elucidate their nonlocal and nonlinear features too.\newline

In practice, nonlocality and nonlinearity can be captured through the two-point susceptibility tensor $\underline{ \bm{\chi }}({\bf r}, {\bf r'})$ of the macroscopic electric polarisation ${\bf P}({\bf r})$ to a change in the electric displacement ${\bf D}({\bf r'})$\cite{Kornyshev1985,hildebrandt2004}:
\begin{equation} 
\frac{\delta {\bf P}(r)}{\delta {\bf D}(r')} \equiv \underline{ \bm{\chi}}( \bm{r}, \bm{r'}). \label{eq0}
\end{equation} In Eq. \eqref{eq0}, nonlocality is expressed through components of $\underline{ \bm{\chi}}( \bm{r}, \bm{r'})$ that differ from a Dirac-delta function while nonlinearity would manifest itself in a functional dependence of these components on the source field $\bf{D}(\bm{r})$.\newline Extensive Molecular Dynamics (MD) simulations with explicit water models such as SPCE or TIPnP \cite{bopp1996} offer an encouraging possibility of probing nonlocal saturated dielectric properties down to the microscopic level but do not furnish analytical expressions of the observables. A radically different approach to consider nonlocal properties of fluid that has also bore fruits to model hydrophobicity is that of Density Functional Theories (DFT) for water  \cite{lum1999,willard2010}. They focus on the description of the density fluctuations at nanometric lengthscale and give a good framework for the study of density fluctuation driven phenomena such as the interaction between hydrophobic objects \cite{chandler2005}. However they do not take into account the molecular and the multipolar nature of the solvent. Mid-way approaches between MD simulations and DFT can be found in Molecular DFT theories or liquid state theory integral equations approches\cite{belloni2018} combined with molecular fields such as  3D-RISM \cite{hirata1981,daCosta12}. Such multi-scale approaches mixing continuum models with insights from classical force fields or MD simulations include the nonlocal properties dielectric properties of water and have been very useful to estimate solvation energies of various kinds of solutes and are still making progress to tackle the proper solvation of activations sites in proteins through combination with electronic DFT methods \cite{Kovalenko17}. These latter mixed approaches have the drawback that, in practice, they tend to put quantities such as the dielectric susceptibility of the solvent --- {\it e.g.} water --- as an input to the theory and not an output of it \cite{jeanmairet2013} and are furthermore not usually analytically tractable. For these reasons, another popular approach is that of implicit solvent continuum models relying on a quadratic Landau-Ginsburg expansion of the free energy with respect to polarisation and density which are able to reproduce qualitative features of the dielectric properties and the structure factor of water \cite{kornyshev1997,maggs2006,berthoumieux2015,berthoumieux2018}.
Nevertheless, these models being linear, they do not include the saturation of the dielectric response  in the presence of strong fields \cite{maggs2006, kornyshev1997}. \newline These nonlinear effects are of prime importance to describe the response of water in the vicinity of charges. Ultimately, the dielectric properties of electrolytes at the nanometric scale are the results of water-water, water-ion and ion-ion interactions and an implicit model treating solvent and ions on the same footing while accounting for nonlocal and nonlinear effects is still missing. Some progress has been made in recent years along these lines \cite{paillusson2010, Blossey14}.

Continuous field-theoretic models derived from a microscopic description of water as a gas of Langevin dipoles give access to the aforementioned saturation effects in water \cite{abrashkin2007,levy2013,adar2018}.These models describe the existence of an hydration shell associated with a low permitivity surrounding  an ion which coherent with the low interfacial permitivity recently measured\cite{fumagalli2018} and can induce the permitivity decrement observed in electrolytes\cite{hasted48}. However the model is purely local and does not include the length scale associated to water-water correlations. 
The goal of this work is therefore to combine the strengths of the nonlocal linear Landau-Ginsburg functionals with that of the nonlinear field-theoretic models by proposing a model for water comprising both nonlocal and nonlinear effects and by probing the dielectric properties of such a medium in the surrounding of an ion.

In this paper, we focus on the dielectric properties of water and propose a nonlocal nonlinear functionnal of the polarization field ${\bf P}$ defined as the density of water dipoles.\\ \\ This paper is organized as follows. In a first part, we present a nonlocal nonlinear functional of the polarization ${\bf P}$ and show that this model captures dielectric properties of bulk water. In a second part we consider the response of this medium to an ion and study the combined effects of nonlocality and nonlinearity. In the third part, we evaluate the full nonlinear nonlocal position-dependent 1-point dielectric response of the polarisation in the hydration shell. The last part is devoted to the conclusion.

\subsection{Description of the model and theory}
We consider a continuous dielectric medium characterised by a statistical polarisation vector field $\bm{ \mathcal{P}}(\bm{r})$ 
and the corresponding electrostatic energy written as a functional of the polarization vector $\bm{ \mathcal{P}}(\bm{r})$ : 
\begin{eqnarray}
\label{bulkhamiltonien}
  \mathcal{H}\lbrack{ \bm{ \mathcal{P}} }\rbrack&=&\frac{1}{2\epsilon_0}\int d^3r\,
\Big [ \gamma(\bm{ \mathcal{P}}(\bm{r})^2+P_0^2)^2
+\kappa_l(\bm{\nabla}\cdot \bm{ \mathcal{P}}(\bm{r}))^2 \nonumber\\
                                  & & + \alpha (\bm{\nabla}
                                     (\bm{\nabla} \cdot \bm{ \mathcal{P}}(\bm{r}))  )^2 \Big ] \nonumber \\
                                && + \frac{1}{2 \epsilon_0}\int d^3r d^3r' \frac{ \bm{\nabla_r} \cdot \bm{ \mathcal{P}}(\bm{r}) \bm{\nabla_{r'} } \cdot \bm{ \mathcal{P}}(\bm{r'})}{4\pi |\bm{r}-\bm{r'}|}. 
\end{eqnarray}
The system being rotational invariant, the appropriate order parameter is $\bm{ \mathcal{P}}(\bm{r})\cdot\bm{ \mathcal{P}}(\bm{r})$. The term $\int d^3r  \gamma(\bm{ \mathcal{P}}(\bm{r})^2+P_0^2)^2$ contains a term scaling in $(\bm{ \mathcal{P}}(\bm{r})\cdot\bm{ \mathcal{P}}(\bm{r}))^2$ corresponding to the first nonlinear contribution to consider\cite{maggs2006}.  The reference polarisation $P_0$ introduces a threshold value for this nonlinear behavior that occurs for $P$ larger than $P_0$. The derivative terms of the functional introduce scale dependent dielectric properties. The last term corresponds to the long-range Coulomb interaction.\\
We then introduce the partition function of the model via:
\begin{equation}
\mathcal{Z}[\bf{D}_0] \equiv \int \mathcal{D}[\bm{\mathcal{P}}] e^{-\beta \mathcal{H}[\bm{\mathcal{P}}]+ \frac{\beta}{\epsilon_0} \int d^3 r \: \bm{ \mathcal{P}}(\bm{r}) \cdot \bf{D_0}(\bm{r}) }, \label{partfunc}
\end{equation}
where $\bf{D}_0(\bm{r})=\epsilon_0{\bf E}_0$ and $E_0$ is the external electrostatic field. We note that $\bf{D}_0(\bm{r})$ also enables $\mathcal{Z}[\bf{D}_0]$ to act as a generating functional of the moments of $\bm{ \mathcal{P}}$. In particular the macroscopic polarisation field reads
\begin{equation}
{\bf P}(\bm{r}) \equiv \langle \bm{ \mathcal{P}}(\bm{r})   \rangle =  \frac{\epsilon_0}{\beta} \frac{1}{\mathcal{Z}[\bf{D}_0]} \frac{\delta \mathcal{Z}[\bf{D}_0]}{\delta \bf{D}_0(\bm{r})},  \label{mean}
\end{equation} and the two-point moment tensor reads
\begin{equation}
\langle  \bm{ \mathcal{P}}(\bm{r}) \otimes \bm{ \mathcal{P}}(\bm{r'}) \rangle = \frac{\epsilon_0^2}{\beta^2} \frac{1}{\mathcal{Z}[{\bf D}_0]} \frac{\delta^2 \mathcal{Z}[\bf{D}_0]}{\delta \bf{D}_0(\bm{r})\delta \bf{D}_0(\bm{r'})}. \label{2moment}
\end{equation}From Eq. \eqref{eq0} it then comes 
\begin{equation}
\underline{\bm{\chi}}(\bm{r},\bm{r'}) = \frac{\beta}{\epsilon_0} \left( \langle  \bm{ \mathcal{P}}(\bm{r})  \otimes \bm{ \mathcal{P}}(\bm{r'}) \rangle - \bf{P}(\bm{r}) \otimes \bf{P}(\bm{r'}) \right). \label{eq1}
\end{equation}

The mean field polarisation ${\bf P}_{MF}(\bm{r})$ of this medium is, by definition, the most probable field configuration {\it i.e.} the field minimising the energy:
\begin{equation}
\left. \frac{\delta \mathcal{H}\lbrack \bm{ \mathcal{P}}\rbrack}{\delta \bm{ \mathcal{P}}(\bm{r})} \right|_{\bm{ \mathcal{P}}(\bm{r})={\bf P}_{MF}(\bm{r})}=\frac{1}{\epsilon_0} {\bf D}_0(\bm{r}).
\end{equation}
and is solution of the equation 
\begin{eqnarray}
&&2 \gamma {\bf P}_{MF}(\bm{r}) \left({\bf P}_{MF}(\bm{r})^2+ P_m^2 \right) - \kappa_l \bm{\nabla} ( \bm{\nabla} \cdot {\bf P}_{MF}(\bm{r})) \nonumber\\ &&  \hspace{5mm} +\alpha \bm{\nabla} ( \bm{\nabla} \cdot (\bm{\nabla} ( \bm{\nabla} \cdot {\bf P}_{MF}(\bm{r})))) = {\bf D}_0(\bm{r}), \label{meanfield}
\end{eqnarray}
with 
\begin{equation}
\label{pm}
P_{m}=\pm P_0\sqrt{1+1/(2\gamma P_0^2)}.
\end{equation}
In bulk, the right hand side of Eq. \eqref{meanfield} is identically zero and the mean polarisation vanishes ${\bf P}_{MF}=0$. The polarisation $P_m$ is a threshold value distinguishing two behaviours of the medium. For polarisation response or fluctuations smaller, respectively larger, than $P_m$, the medium behaves as a linear medium or a nonlinear medium. 

\subsection{Susceptibility and Bulk polarisation correlations of water to the Gaussian expansion}
In order to capture the susceptibility tensor $\underline{\bm{\chi}}$,
%In order to capture the dipolar correlations in this medium, one has to go beyond the mean-field description. We start by defining the partition function $\mathcal{Z}$ of the system and introduce an auxiliary field ${\bf h}(r)$, such that
%\begin{equation}
%\label{partfunc}
%\mathcal{Z}[{\bf h}]=\int \mathcal{D}[{\bf P}] e^{-\beta \mathcal{H}[{\bf P}]-\int d^3r {\bf h}(r){\bf P}(r)}.
%\end{equation}
%and express the polarization correlations as
%\begin{equation}
%\label{correl}
%\langle{\bf  P}(r) {\bf P}(r')\rangle = \frac{\delta^2 \mathcal{Z}[{\bf h}]}{\beta^2 \mathcal{Z}[{\bf h}]\delta {\bf h}_r\delta {\bf h}_{r'}}.
%\end{equation}
the partition function in Eq. (\ref{partfunc}) can be estimated by expanding $\mathcal{H}[\bm{\mathcal{P}}]$ about the bulk mean field solution ${\bf P}_{MF}(r)=0$,
\begin{eqnarray}
& &\mathcal{Z}[{\bf D}_0]\approx e^{-\beta \mathcal{H} [{\bf P}_{MF}(\bm{r})=0]}\nonumber\\
&\times& \int\mathcal{D}[\delta {\bf P}]e^{-\frac{\beta}{2}\int d^3r \int d^3r'\delta {\bf P}(\bm{r}) \underline{{\bf M}}^{-1}_{\bm{r},\bm{r'}} \delta  {\bf P}(\bm{r'})+ \frac{\beta}{\epsilon_0} \int d^3r {\bf D}_0(\bm{r}) \cdot {\bf P}(\bm{r})} \nonumber \\
&\approx &\sqrt{\frac{2\pi}{ \beta {\rm det}(\underline{{\bf M}}^{-1}) }} e^{-\beta \mathcal{H} [{\bf P}_{MF}(\bm{r})=0]} e^{\frac{\beta}{2 \epsilon_0^2}\int d^3r \int d^3 r'  {\bf D}_0(\bm{r}) \underline{{\bf M}}_{\bm{r},\bm{r'}}  {\bf D}_0(\bm{r'})} \label{eq3}
\end{eqnarray}
with $ \bm{\mathcal{P}}(\bm{r})={\bf P}_{MF}+\delta {\bf P}( \bm{r})$ and $ \underline{{\bf M}}^{-1}_{r,r'}$ being the second functional derivative of $\mathcal{H}[\bm{\mathcal{P}}]$ with respect to $\bm{\mathcal{P}}$ and evaluated at the bulk mean field value $\left.\underline{{\bf M}}^{-1}_{r,r'} \equiv \frac{\delta \mathcal{H}}{\delta \bm{\mathcal{P}}(\bm{r}) \delta  \bm{\mathcal{P}}(\bm{r'})}\right|_{P_{MF}=0}$. Using Eq. (\ref{bulkhamiltonien}) one finds
\begin{eqnarray}
\underline{{\bf M}}^{-1}_{r,r'}&=&\frac{1}{\epsilon_0}\Big(2\gamma P_m^2{\bf Id}+\kappa_l ( \bm{\nabla_r} \cdot)\otimes( \bm{\nabla_r} \cdot)\nonumber\\ &+&\alpha( \bm{\nabla_r} \bm{\nabla_r }\cdot)\otimes( \bm{\nabla_r} \bm{\nabla_r} \cdot)\Big)\delta(\bm{r}- \bm{r'}).
\end{eqnarray}
Applying Eqs. \eqref{mean} and \eqref{eq1} to Eq.\eqref{eq3} it follows that at the gaussian level we have:
\begin{eqnarray}
&& {\bf P}(\bm{r}) \overset{Gauss.}{=} \frac{1}{\epsilon_0}\int d^3 r' \: \underline{{\bf M}}_{\bm{r},\bm{r'}} {\bf D}_0(\bm{r'}), \label{gausspol} \\
&& \underline{\bm{\chi}}(\bm{r},\bm{r'}) \overset{Gauss.}{=} \frac{1}{\epsilon_0} \underline{{\bf M}}_{\bm{r},\bm{r'}}  .\label{gausssuscept}
\end{eqnarray}From Eq. \eqref{gausspol}, we retrieve that the response of the mean polarisation to an external field at the gaussian level of approximation is linear, albeit potentially nonlocal.\\

%The dipole-dipole correlations are equal, in a Gaussian approximation, to
%\begin{eqnarray}
%\langle P(r) P(r')\rangle =\frac{\epsilon_0}{2\beta} {\bf \chi}_{r,r'}
%\end{eqnarray}
%with $\chi_{r,r'}=\epsilon_0 {\bf M}_{r,r'}$.

To resolve for $\underline{\bm{\chi}}$ we make use of the fact that the tensor components of $\underline{{\bf M}}^{-1}$ take a simple form in Fourier space decomposable in terms of longitudinal and transverse parts:
\begin{eqnarray}
\label{Mm1}
& & M^{-1}_{ij}(q)=M^{-1}_{\perp}(q)\left(\delta_{ij}-\frac{q_iq_j}{q^2}\right)+ M^{-1}_{\parallel}(q)\frac{q_iq_j}{q^2}\nonumber\\
&{\rm with}& \quad M^{-1}_{\perp}(q)=2\gamma P_m^2/\epsilon_0,\\
&{\rm and}&  \quad M^{-1}_{\parallel}(q)=(2\gamma P_m^2+\kappa_l q^2+\alpha q^4)/\epsilon_0,
 \end{eqnarray}where $\bm{q}$ is the wave vector and $q$ its magnitude.\\
 Using Eqs. (\ref{2moment},\ref{eq3},\ref{gausssuscept},\ref{Mm1}), the bulk polarization correlations calculated to the Gaussian order can be expressed as
\begin{equation}
\label{pcorelGauss}
\langle P_{\alpha}(0)P_{\beta}(r)\rangle=\epsilon_0 k_BT \int d^3q e^{i{\bf q} \cdot {\bf r}}\underline{\bm{\chi(q)}}_{\alpha\beta}.
\end{equation}

The longitudinal susceptibility associated to this model at the Gaussian level, 
\begin{equation}
\label{chiq}
\chi_{_{\parallel \: \rm lin}} (q) = \frac{1}{2\gamma P_m^2+\kappa_l q^2+\alpha q^4},
\end{equation}
has been shown to qualitatively reproduce the susceptibility of water obtained both experimentally and using explicit models of the liquid such as SPC/E\cite{jeanmairet2013}. 
In particular, it can fit the macroscopic value of the susceptibility and the position and value of its main maximum.  
The expression of the polarization correlation obtained from Eq. (\ref{pcorelGauss}) contains an oscillating exponentially short range term that adds to and dominates the long range Coulombic term in $1/r^3$ on nanometric distances \cite{berthoumieux2015}. Note that in this work, the transverse susceptibility $\chi_{\perp}(q)=1/2\gamma P_0^2$ is supposed to be local, however the nonlocal transverse effects can be considered in the present framework \cite{berthoumieux2015}.

\subsection{Quantifying nonlinearities in bulk}

We now want to characterise the nonlinear effects, if any, owing to the Hamiltonian in Eq. (\ref{bulkhamiltonien}). Following Ref. \cite{kornyshevsat1997}, we note that in the linear and homogeneous case ${\bf P}(\bm{r}) = \int d^3 r' \: \underline{\bm{\mathcal{\chi}}}_{_{\rm lin}}(\bm{r} - \bm{r'}) {\bf D}_0(\bm{r'})$. Thus, if the external field ${\bf D}_0$ is of the form ${\bf D}_0(\bm{r}) = D_0 \cos(q x) {\bf u}_x$, with $x$ a Cartesian coordinate and ${\bf u}_x$ the associated unit vector, we have a polarisation of the form ${\bf P}=P(x) {\bf u}_x$ satisfying
\begin{eqnarray}
P(x) &= & \int dx' \: \chi_{_{\parallel \: \rm lin}}(x-x') D_0 \cos(qx') \nonumber \\
 &=& \chi_{_{\parallel \: \rm lin}}(q) D_0 \cos(q x), \label{basicidea}
\end{eqnarray} the transverse part of the response vanishing in this case. From Eq. \eqref{basicidea} we see that, provided the system is probed with an external field of the proper form, one can write $\chi_{_{\parallel \: \rm lin}}(q) = \frac{\partial P(0)}{\partial D_0} $. We generalise this result by defining the Fourier transform of an effective longitudinal susceptibility $\chi_{_{\parallel \: eff}}$ associated to a general polarisation ${\bf P}[D_0 \cos(qx)]$:
\begin{equation}
\chi_{_{\parallel \: eff}}(q, D_0) \equiv \left. \frac{\partial P(q, D_0, x) }{\partial D_0} \right|_{x=0} \label{chieff}
\end{equation}

%we determine the  response of the dielectric medium to a small oscillating displacement field ${\bf D}(x)=D_0 \cos(q x){\bf u}_x$, where $q$ is the spacial frequency, $x$ a Cartesian coordinate and ${\bf u}_x$ is the unitary vector associated.
%The Hamiltonian of the perturbed system can be written as 
%\begin{equation}
%\label{hamiltonianfreq}
%\mathcal{H}_{pert}[{\bf P}]=\mathcal{H}[{\bf P}]-\frac{1}{\epsilon_0}\int d^3r {\bf P}(r) {\bf D}(r).
%\end{equation}
At the mean field level, in presence of the external perturbation, the amplitude $P(x)$ is solution of the equation
\begin{equation}
P(x)^3+P_m^2\left(P(x)-\chi_b D(x)\right)-\frac{\kappa_l}{2\gamma}P^{(2)}(x) +\frac{\alpha}{2 \gamma}P^{(4)}(x)=0, \label{1dmeanfield}
\end{equation}
obtained from Eq. \eqref{meanfield} and where $P^{(i)}(x)$ stands for the $i^{th}$ derivative of $P(x)$ and with $\chi_b \equiv \chi_{_{\parallel \: lin}}(0)$ the linear macroscopic longitudinal susceptibility of the medium given in Eq. (\ref{chiq}). $P(x)$ can be expressed as an expansion in $D_0$  to the third order, such that ({\it cf.} Appendix C) $P(x)=D_0 \chi_{_{\parallel \: lin}}(q) \cos(qx)-\frac{D_0^3}{4P_m^2\chi_b}\left(3\chi_{_{\parallel \: lin}}(q)^4\cos(q x)+\chi_{_{\parallel \: lin}}(q)^3\chi_{_{\parallel \: lin}}(3q) \cos(3 q x)\right)$. The negative sign before the term in $D_0^3$ reproduces the saturation effect expected from a Hamiltonian in $P^4$. 
Note that the term in $D_0^3$ has to be much smaller than the term in $D_0$ to ensure the validity of the performed expansion, translating into the condition $D_0 \ll \frac{2 P_m}{\chi_{_{\parallel \: lin}}(q)\left(3 \chi_{_{\parallel \: lin}}(q)/\chi_b+\chi_{_{\parallel \: lin}}(3q)/\chi_b\right)^{1/2}}$. 
The corresponding effective susceptibility can readily be obtained from Eq. \eqref{chieff} and  is plotted in Fig. 1 with a comparison with the gaussian susceptibility given in Eq. (\ref{chiq}.) The model reproduces qualitatively the results obtained with molecular dynamics simulations by Kornyshev and coworkers with the BJH model of water \cite{bopp1983}. The saturation effects occur first for the wave number associated with the maximum, the susceptibility is flattened in Fourier space, reducing the nonlocal character of the medium and this effect increases with the amplitude of as illustrated in Fig. 1. a. The Figure 1. b. shows the comparison of simulations results obtained with the explicit model of water SPC/E \cite{yeh1999} and the theoretical model presented here for the dielectric permitivity of water under constant field,
\begin{eqnarray}
\label{epsilonE}
& &\epsilon_r(E_0)=1/(1-\chi_{eff}(0,\epsilon_0 E_0)) \quad {\rm with} \\
& &\chi_{eff}(q,D_0)\approx\chi_{\parallel}(q)-\frac{3}{2} D_0^2 \gamma \chi_{\parallel}(q)^3 \left(3\chi_{\parallel}(q)+\chi_{\parallel}(3 q)\right) \nonumber
\end{eqnarray}

calculated using Eq. (\ref{chieff}) expanded to $D_0^2$ with different values of $P_m$. One sees that the model reproduces the trend of the simulation results for $P_m\approx 0.09 C.\AA^{-1}$. For large excitation fields, the second order expansion of $\epsilon_r(E_0)$ plotted here fails to reproduce the simulation results.

\begin{figure}
\includegraphics[scale=0.3]{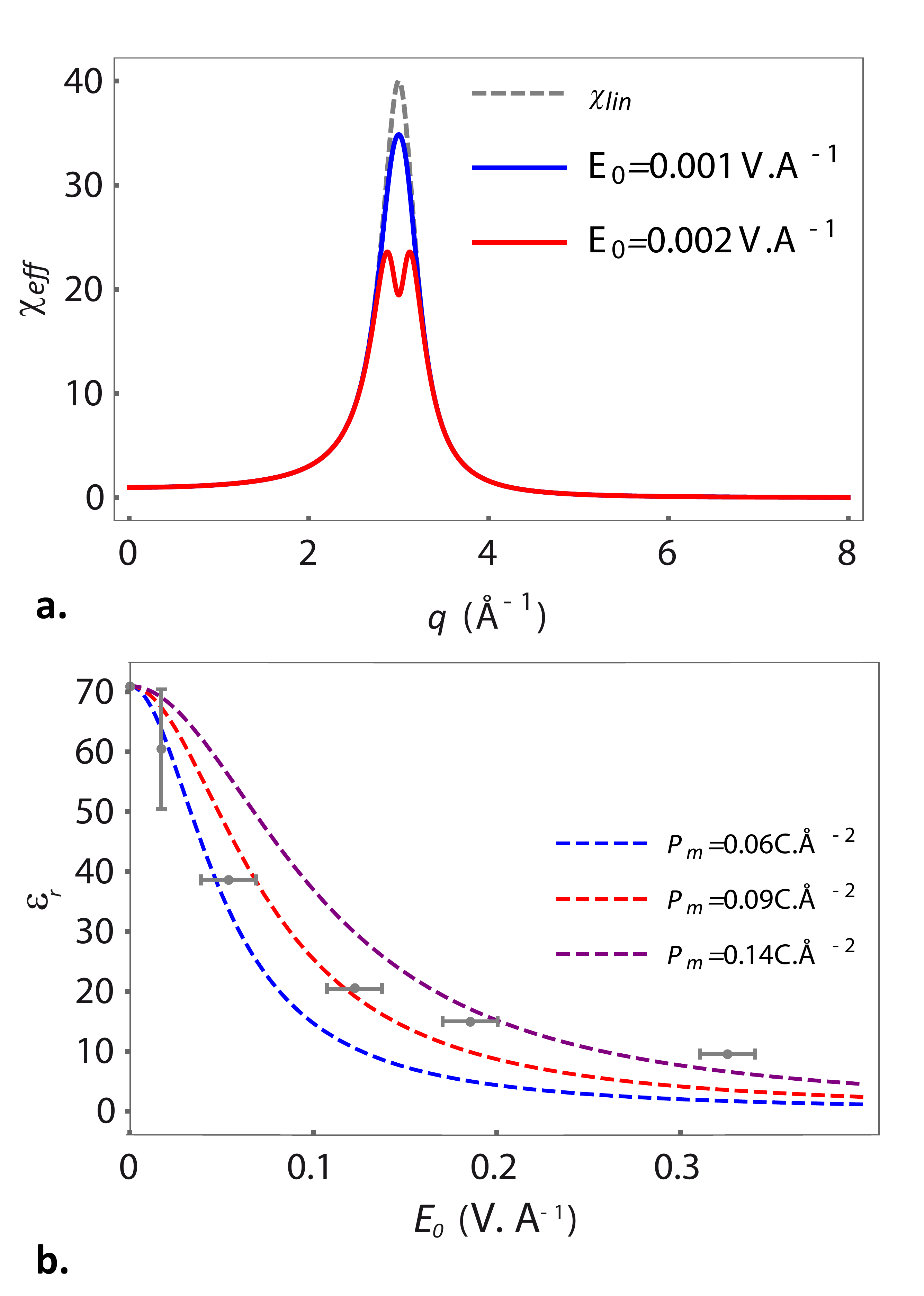}
\caption{{\bf a.} Susceptibility and effective response of a non linear dielectric medium. The dashed plot represents the susceptibility of the medium (Eq. (\ref{chiq})). The blue and the red curves represent the effective response given in Eq. (\ref{chieff}) of a medium submitted to a small excitation, $E_0$, for $P_m=0.09 C.\AA^{-2}$. The curves are obtained for $2\gamma P_m^2= 1.01$, $\kappa_l=-0.22$ \AA$^{2}$, $\alpha=0.012$ \AA$^{4}$. {\bf b.} Dielectric permitivity under constant field,  $\epsilon_r(E_0)$ (Eq. (\ref{epsilonE})), for different values of $P_m$ compared to results obtained with Molecular Dynamics  using SPC/E model for water (grey points and error bars).   The molecular dynamics results are reproduced from the Fig. 11 of Yeh I.-C. et al., J. Chem. Phys. {\bf 110} 7935-7942, 1999.  The curves are obtained for $2\gamma P_m^2= 1.01$, $\kappa_l=-0.22$ \AA$^{2}$, $\alpha=0.012$ \AA$^{4}$.}
\end{figure}

\section{Position-dependent dielectric response induced by a point-like ion}
\subsection{Polarisation around an ion for a nonlocal nonlinear medium}
We now consider the response of the medium to the electrostatic field generated by an ion located in $r=0$. The %Hamiltonian of the perturbed system can be written as 
%\begin{equation}
%\label{hamiltonianion}
%\mathcal{H}_{ion}[{\bf P}]=\mathcal{H}[{\bf P}]-\frac{1}{\epsilon_0}\int d^3r {\bf P}(r) {\bf D}_{ion}(r),
%\end{equation}
%where 
corresponding electric displacement field reads ${\bf D}_{ion}(r)=D_{ion}(r){\bf u}_r$ with 
\begin{equation}
D_{ion}(r)=\frac{e}{4 \pi r^2},
\end{equation}
$e$ being the charge of the ion. At the mean field level, the macroscopic polarisation field of the form ${\bf P}_{MF}(\bm{r}) = P_i(r) {\bf u}_r$ satisfies Eq. \eqref{meanfield} which can be written in the following dimensionless form: 
\begin{eqnarray}
\label{eqion}
& &\left(\frac{P_i(r)}{P_m}\right)^3+\frac{P_i(r)}{P_m}-\kappa_l\chi_b\left(\frac{P_i''(r)}{P_m}+2\frac{P_i'(r)}{r P_m}-2\frac{P_i(r)}{r^2P_m}\right)\nonumber\\&+&\alpha\chi_b \left(\frac{P_i^{(4)}(r)}{P_m}+4\frac{P_i^{(3)}}{r P_m}-4\frac{P''_i(r)}{r^2 P_m}\right)=\frac{l^2}{r^2}.
\end{eqnarray}
The distance,
\begin{equation} 
\label{l}
l=\sqrt{\frac{e}{4 \pi P_m}\chi_b},
\end{equation} 
characterises the range of the saturation effects for $P_i(r)$ as it is the distance beyond which the local linear response $P_{ll}=\chi_b e/4\pi l^2$ to the electrostatic field generated by the ion is smaller than $P_m$.

The response of a local nonlinear dielectric medium - obtained by neglecting all the spacial derivatives of $P$ in Eq. (\ref{eqion}) - and of a nonlocal linear dielectric medium - obtained by neglecting the term in $P^3/P_m^3$ in Eq (\ref{eqion}) - have been studied in the literature \cite{levy2013,berthoumieux2018} and some relevant results are presented in Appendix A and B.

We compare the polarisation around an ion obtained in this framework with the molecular dynamics simulation results. We solve numerically the complete equation given in Eq. (\ref{eqion}) with the Mathematica software. To do so, we fix the value of the polarisation and its derivative equal to the one of the first hydration shell around a chloride obtained with molecular dynamics simulations (\cite{jeanmairet2016}) $P_i$(3.16 \AA)= -0.68 C.\AA$^{-2}$ and $P_i'$(3.16\AA)= 1.8 C.\AA$^{-3}$, and at a distance of 1 nm we impose local linear conditions, {\it i. e.} $P_i(10)=P_{loc}(10)$ and $P'_i(10)=P'_{loc}(10)$
with $P_{loc}(r)=\chi_b e/4 \pi r^2$ and $e=-1.6$ $10^{-19}$C for the chloride.

The results are presented in the Figure 2. 
The Fig. 2 {\bf a.} represents the results obtained with molecular dynamics simulations (red points ) with a SPC/E model for water and a charged Lennard-Jones sphere for the chloride (see \cite{jeanmairet2016} for all the parameters), the local linear polarisation $P_{loc}(r)=\chi_b e/4 \pi r^2$ (dashed gray line) and the solution of Eq. (\ref{eqion}) for $P_m=0.1$ C.\AA$^{-2}$, i. e. $l$=3.5 \AA. The polarisation presents a oscillating decay over 1 nm before converging to the local polarisation $P_{loc}(r)$. The model presented here reproduces the both the oscillations and the range of the decay of the nonlocal contribution. This qualitative agreement with the molecular dynamics results is encouraging.

The figure 2 {\bf b} represent solution of the equation Eq. (\ref{eqion})  for the aformentionned bondary conditions and for increasing values of $l$. One sees that the range of the oscillation regime increases with the decrease of $l$.

\vspace{2mm}

Our nonlocal nonlinear dielectric model can thus reproduce both the reported saturation and oscillations behaviour of the polarisation response of water in the neighbourhood of an ion as obtained from MD simulations. The range of the perturbation of a charged inclusion on dielectric properties of water is about 1 nm as illustrated by computation of the polarization or of the dielectric permitivity around a solute reproduced in Fig. 2. {\bf a}. \cite{jeanmairet2016}. At first glance, a hydration length 0.3 nm$< l<$0.4 nm for a monovalent ion corresponding to a saturation polarisation 0.08 C.\AA$^{-2}<P_m<$ 0.14 \AA$^{-2}$ gives results comparable with the qualitative behaviour of water reported from Ref.  \cite{shaff2015,jeanmairet2016,jeanmairet2018}. More detailed MD simulations could permit to determine the best fitting value of $P_m$ for the water response to an ion. However, to our knowledge, the polarisation response of water to an ion beyond the first hydration shell has not yet been computed with MD. 
 
\begin{figure}
\includegraphics[scale=0.4]{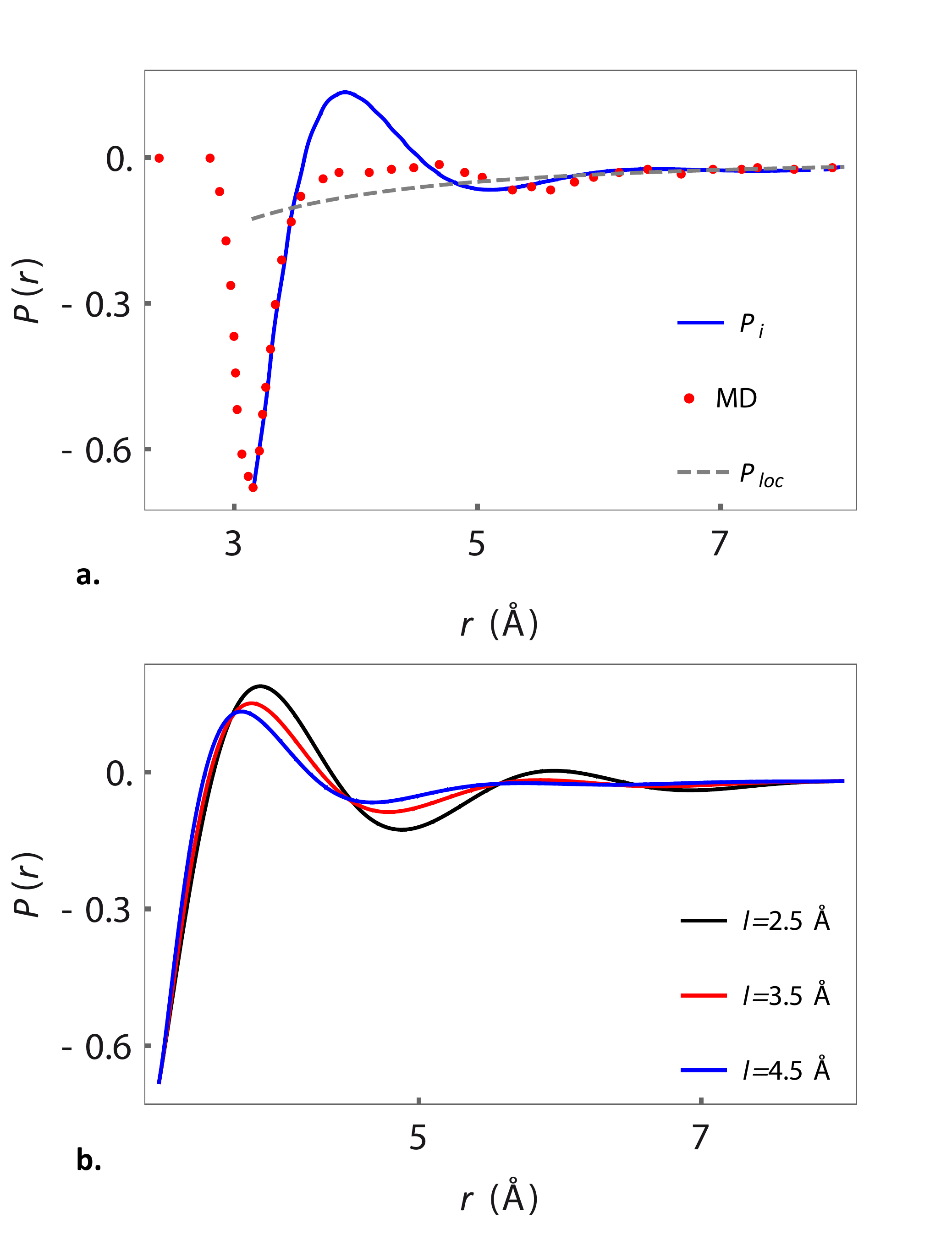}
\caption{Polarisation response $P_i(r)$ (C.\AA$^{-2}$) to the electrostatic field generated by a monovalent anion. ({\bf a.} The plain curve represents $P_i(r)$, solution of Eq. (\ref{eqion}), as a function of $r$, for the set of parameters given in Fig. 1 and $P_m=$ 0.1 C. \AA$^{-2}$, $l=0.35$ nm. The local polarisation $P_{loc}$ is represented by the gray dashed line and the molecular dynamics results are indicated with the red points, reproduced from the Fig 11. of Jeanmairet, G. et al., J. Phys: Condens. Matter {\bf 28}. (2016) 244005. ({ \bf b.}). Polarisation response  $P_i(r)$ (C.\AA$^{-2}$) , solution of Eq. (\ref{eqion}), as a function of $r$, for increasing values of $l$.}
\end{figure}

\subsection{Dielectric susceptibility in the vicinity of the ion}
We now look at the macroscopic one-point dielectric susceptibility of water in the vicinity of the ion.
The dielectric longitudinal susceptibility is not homogeneous in a solution and depends in particular on the local concentration of ions. The static dielectric constant of an electrolyte is known to be a decreasing function of the salt concentration\cite{hasted48}. The intuitive comprehension of this phenomenon is that each ion generates a strong field that 'freezes' the dipoles around it, which can not respond anymore to an external excitation\cite{gavish16}.

A local nonlinear dielectric medium can present a vanishing dielectric permitivity in the vicinity of an ion as shown by Orland and coworkers \cite{levy2013}. Following their approach we determine here the dielectric susceptibility of a nonlocal nonlinear medium in the hydration shell of an ion pinned at $\bm{r}=0$. 
 
To do so we calculate the mean-field response of the medium ${\bf P}_{D_0} $ to an additional source field ${\bf D}_0$ supplementing that of the fixed ion. Using Eq. \eqref{meanfield}, it follows:
\begin{eqnarray}
\label{eqionD0}
&& \frac{{\bf P}_{D_0}^3(\bm{r})}{P_m^3}+\frac{{\bf P}_{D_0}(\bm{r})}{P_m}+\kappa_l \bm{\nabla} \left( \bm{\nabla} \cdot \frac{{\bf P}_{D_0}(\bm{r})}{P_m} \right) \nonumber \\ && \hspace{2mm} +\alpha \bm{\nabla} \left( \bm{\nabla} \cdot  \left( \bm{\nabla} \left( \bm{\nabla} \cdot \frac{{\bf P}_{D_0}(\bm{r})}{P_m}\right) \right) \right) \nonumber\\ &&\hspace{4mm} = \frac{l^2}{r^2}{\bf u}_r+\chi_b\frac{{\bf D}_0}{P_m}.
\end{eqnarray}
%The susceptibility tensor components are given in spherical coordinates by 
%\begin{equation}
%\chi_{i, ij}({\bf r})=\frac{\partial P_{D,i}(\bf{r})}{\partial D_{0,j}(\bf{r})}|_{D0=0}, \quad (i,j)=(r,\theta,\phi).
%\end{equation}
From Eq. \eqref{chieff}, the bulk static longitudinal susceptibility can be obtained by finding the polarisation response to a directional uniform displacement field. In this part we consider a perturbation to the existing field generated by the ion of the form ${\bf D}_0 =D_0{\bf u}_r$. We can probe the distance-dependent effective susceptibility around the ion for a radial polarisation response ${\bf P}_{D_0}(\bm{r}) = P_{D_0}(r) {\bf u}_r $,  via
\begin{equation}
\label{chi}
\chi_{ion}(r)= \left.\frac{\partial P_{D_0}(r)}{\partial D_{0}} \right|_{D_0=0}
\end{equation}
%which is the only non vanishing for a local nonlinear medium. 
Eq.\eqref{chi} is evaluated numerically by solving Eq. (\ref{eqionD0}) for a radial excitation field and a small value of $D_0$ as $\lim_{D0 \rightarrow 0} (P(r,D_0)-P(r,0))/D_0$ and the result is plotted in Fig. 3. 

The figure 3 presents the dielectric susceptibility in the vicinity of a monovalent ion for a nonlocal nonlinear medium, the plain blue curve, for a nonlinear local medium, the gray dotted line, and for a linear nonlocal medium, the red dashed line. For $r\ll l$, $\chi_{ion}(r)$ behaves as the susceptibility of a nonlinear local medium. The ion is surrounded by a zone of vanishing susceptibility indicating that the solvent molecules located in the first hydration shell of the ion are 'frozen' {\it i. e.} that they interact electrostatically only with the ion and do not respond to an external field. For a nonlinear local medium, as one moves away from the ion, the molecules see an homogeneous medium associated with a given bulk dielectric susceptibility. For a nonlocal nonlinear medium, we observe an oscillating behaviour of over-responses and under-responses around the bulk value due to the nonlocality of the medium. The influence of the ion extends over few nanometers for this set of parameters.

\begin{figure}
\includegraphics[scale=0.4]{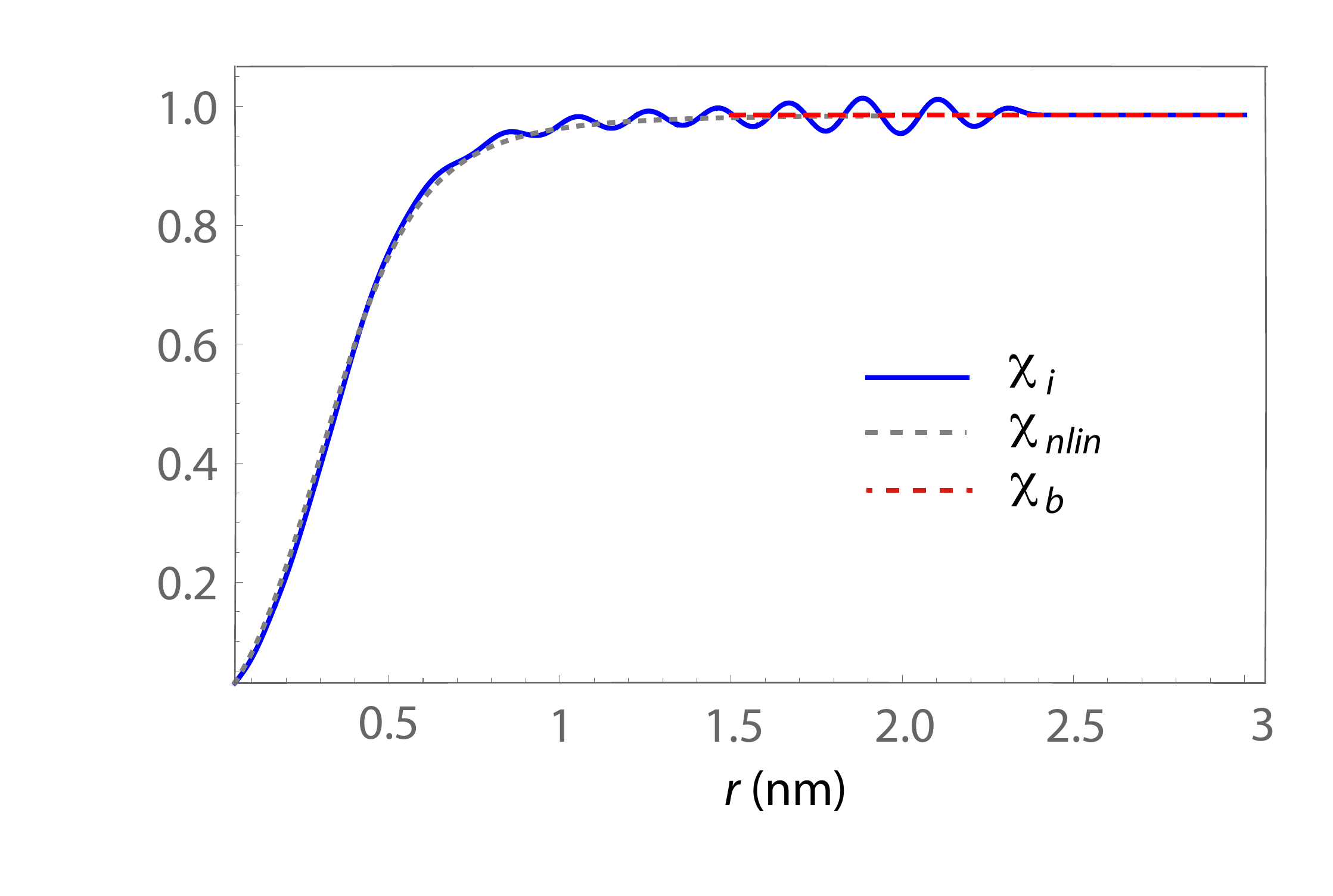}
\caption{Susceptibility for a nonlinear nonlocal medium around an ion. The blue solid line represents $\chi_{i}(r)$ (relative to the bulk value) given in Eq. (\ref{chi}) for a monovalent ion and $P_m=$ 0.31 C. \AA$^{-2}$, $l=0.2$ nm. The dotted line represents the susceptibility around an ion for a nonlinear local medium given in Eq. (\ref{chinlin}) and the red dashed line represents the bulk susceptibility, $\chi_b \equiv \chi_{_{\parallel \: lin}}(q=0)$. }
\end{figure}

\section{Conclusion}

In this article, we have proposed a functional of the polarisation {\bf P} including non Gaussian and nonlocal terms and reproducing the dielectric properties of water at the nanometric scale. We have determined the position-dependent polarisation around a point-like ion obtained and showed the existence of a hydration shell whose range is characterised both by a saturation length $l$ corresponding to a distance to which the field generated by the ion is equal to the saturation polarisation of the medium and a nonlocal length $\lambda_e$ owing to the dipolar correlations in bulk water. We have shown that one effect of nonlocality is to extend the range of saturation effects over length-scales corresponding to short-range molecular interactions. The behaviour obtained for the polarisation field around an ion is in qualitative agreement with molecular dynamics results and a comparison with more detailed results would permit to refine the parameter values to describe water. Finally, we determined the macroscopic susceptibility around an ion and found layers of substantial over- and under- screening around it, already observed in water mediated electrostatic interactions \cite{rottler2009}.

To conclude, this work demonstrates the potential of Landau-Ginsburg approaches extended beyond Gaussian and local models and that semimicroscopic models of water, such as the functional in $P^4$ proposed here, are a good compromise between calculation feasibility and molecular details and could be used to efficiently model the dielectric properties of electrolytes. 

\vspace{2mm}
\textbf{Acknowledgements:} HB acknowledges support from the CNRS through the D\'efi INFINITI - AAP 2018.

\section*{Appendix}

\subsection{Response of a local nonlinear dielectric medium to the electrostatic field generated by an ion}
In this section, we determine the response ${\bf P}_{nlin}(\bm{r})=P_{nlin}(r){\bf u}_r$ of a local nonlinear dielectric medium to a field generated by an ion located at $r=0$. The amplitude of the response is solution of the equation,
\begin{equation}
\tag{A1}
\label{eqnonlinearloc}
(P_{nlin}(r))^3+P_m^2(P_{nlin}(r)-P_1(r))=0
\end{equation}
obtained from Eq. (\ref{eqion}) by neglecting the nonlocal terms. $P_{1}(r)$ is equal to $\chi_bD_{ion}(r)$.
This Eq. (\ref{eqnonlinearloc}) has been previously derived and solved by Orland and coworkers in the framework of the Poisson-Dipolar model they developed to model aqueous electrolytes \cite{levy2012,levy2013}. Following their work, we express $P_{nlin}(r)$ as the root of this third-order polynomial equation,
\begin{equation}
\tag{A2}
\label{Pnlin}
%P_{nlin}(r)&=&P_m\left(\frac{P_1(r)}{2 P_m}+\sqrt{\frac{P_1(r)^2}{4 P_m^2}+\frac{1}{27}}\right)^{1/3}\nonumber\\&+&P_m\left(\frac{P_1(r)}{2 P_m}-\sqrt{\frac{P_1(r)^2}{4 P_m^2}+\frac{1}{27}}\right)^{1/3} \\
P_{nlin}(r)=P_m f \left( \frac{l^2}{r^2}\right), 
\end{equation}
with
$\l=\sqrt{\frac{e}{4 \pi P_m}\chi_b}$ and $f(x)=\left(x/2+\sqrt{x^2/4+1/27}\right)^{1/3}+\left(x/2-\sqrt{x^2/4+1/27}\right)^{1/3}$.
The length $l$ defines two regions: the hydration shell $r \ll l$ in which the polarisation response is attenuated compared to the linear regimes and where the the dipoles of the solvent are frozen by the electrostatic field and a region far from the ion $r \gg l$  where the dipoles respond linearly to the field. Approximate expressions of $P_{nlin}(r)$ are easily obtained in these two regions 
\begin{align}
P_{nlin}(r)&\approx  P_m\left(\frac{l^2}{r^2}\right)^{1/3}, \quad {\rm for} \quad r \ll l  \tag{A3 } \label{Pnlinclose} \\
P_{nlin}(r)&\approx  \chi_d D_{ion}(r), \quad {\rm for} \quad r \gg l. \tag{A4}  \label{Pnlinfar}
\end{align}
The response $P_{nlin}(r)$ is plotted in Fig. 4 for increasing values of  $l$ and compared to the response a linear medium. One sees the saturation effect for $r \le l$ and the rapid convergence to a linear response for $r \ge l$. Molecular dynamics results \cite{jeanmairet2016, shaff2015} have shown that a local linear dielectric response is obtained around monovalent ions for a distance of about 1 nm  which leads to $l$ comprise between 2 \AA and 4 \AA as illustrated in Fig. 4.
\begin{figure}[h]
\includegraphics[scale=0.45]{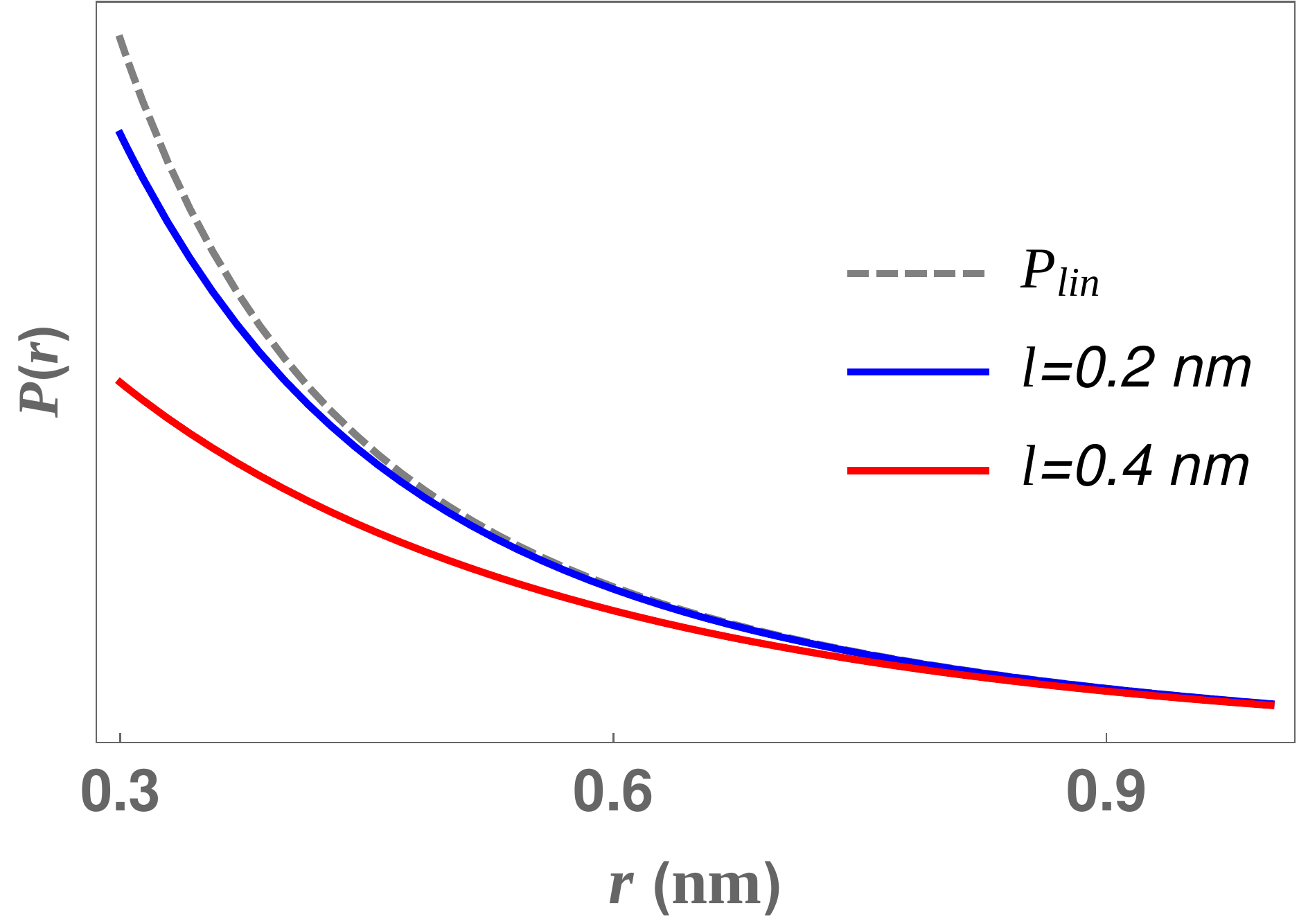}
\caption{ Responses of linear and non linear dielectric media to the electrostatic field generated by a monovalent ion. The dashed plot represents the response of a linear medium, $P_{lin}(r)=\chi_b D(r)$, with $\chi_b=71/70$, the blue and red curves correspond to the response $P_{nlin}(r)$ of nonlinear media associated with increasing hydration length $l$=0.2 nm and $l$=0.4 nm.   The polarisation $P$ is expressed in C. \AA$^{-2}$}
\end{figure}

The macroscopic susceptibility around the ion is obtained by calculating the response of the polarisation in the presence of the ion to a constant field ${\bf D}_0$.
We first determine the expression of the polarisation ${\bf P}_{nlin,D}$ solution of the equation
\begin{equation}
\tag{A5}
\label{eqnonlinsusc}
{ P}_{nlin}(\bm{r})^2 {\bf P}_{nlin}(\bm{r}) +P_m^2( {\bf P}_{nlin}(\bm{r})-\chi_b({\bf D}_{ion}(r)+{\bf D}_{0}) )=0.
\end{equation}
One sees that ${\bf P}_{nlin}(r)$ is collinear to $({\bf D}_{ion}(r)+{\bf D}_{0})$ in all points $r$ and that its amplitude $P_{nlin,D}(r)$, is equal to $P_{nlin,D}(r)=f(l^2/r^2+D_0\chi_b/P_m)$ if ${\bf D}_{0}$ is radial.
Defining the nonlinear position-dependent susceptibility in the presence of the ion as
\begin{equation}
\tag{A6}
    \chi_{nlin}(r)=\frac{\partial P_{nlin,D}(r)}{\partial D_{0}}|_{D_0=0}. \label{chinlin}
\end{equation}
This function is plotted on Fig. 3.

\subsection{Response of nonlocal linear medium to an electrostatic field generated by an ion}
In this section, we calculate the response $P_{nloc}(r)$ of a nonlocal linear medium to the electrostatic field generated by an ion.

Far from the ion ($r \gg l$) the polarisation tends to zero and one can linearise the Eq. (\ref{eqion}) by neglecting the nonlinear term in $(P/P_m)^3$. One obtains the following linear differential equation for a polarisation of the form ${\bf P}_{nloc}(\bm{r})=P_{nloc}(r){\bf u}_r$
\begin{align}
& \frac{P_{nloc}(r)}{P_m}-\kappa_l\chi_b\left(\frac{P_{nloc}''(r)}{P_m}+2\frac{P_{nloc}'(r)}{r P_m}-2\frac{P_{nloc}(r)}{r^2P_m}\right) \nonumber\\&\:\:\:\: +\alpha\chi_b \left(\frac{P_{nloc}^{(4)}(r)}{P_m}+4\frac{P_{nloc}^{(3)}}{r P_m}-4\frac{P''_{nloc}(r)}{r^2 P_m}\right)=\frac{l^2}{r^2}. \tag{B1}
\end{align}
The solution of this equation has been already studied in a previous work \cite{berthoumieux2018} and we only give the main steps to obtain it. Using the potential $\psi(r)$ such as $P(r)= d\psi(r)/dr$ and  writing the linearised equation in Fourier space, one obtains
\begin{align}
{\tilde \psi}(q)&= \frac{e}{q^2(2\gamma P_m^2+\kappa_l q^2+\alpha q^4},\\
&= \frac{e}{q^2}\chi_{_{\parallel \: lin}}(q). \tag{B2}
\end{align}
The expression of the potentiel $\psi(r)$ in real space is then 

\begin{align}
\psi(r)&= -\chi_b\frac{e}{4\pi r}+\frac{1}{4\pi r\alpha}\frac{\lambda_e\lambda_o}{\left(\frac{1}{\lambda_o^2}-\frac{1}{\lambda_e^2}\right)^2+\frac{2}{\lambda_e^2\lambda^2_o}}
e^{-r/\lambda_e} \nonumber \\   
 &\times  \left(\frac{2}{\lambda_e^2\lambda^2_o}\cos(r/\lambda_o)+\left(\frac{1}{\lambda_o^2}-\frac{1}{\lambda_e^2}\right)\sin(r/\lambda_o)\right) \tag{B3} \label{psiloc}
\end{align}
where $\lambda_e$ and $\lambda_o$ are the decay and the oscillation length of the dielectric response of the medium and are equal to
\begin{equation}
\tag{B4}
\label{lambda}
\lambda_e=\frac{\sqrt{2}}{q_0\sqrt{1/\sqrt{\zeta}-1}}, \quad \lambda_o=\frac{\sqrt{2}}{q_0\sqrt{1/\sqrt{\zeta}+1}},
\end{equation}
with $q_0^2=\kappa_l/2\alpha$ and $\zeta=\alpha q_0^4 \chi_b$. 

The linear nonlocal polarisation vector field ${\bf P}_{nloc}$ owing to the ion is then equal to
\begin{equation}
\tag{B5}
\label{Pnloc}
{\bf P}_{nloc}(r)=\overrightarrow{\nabla}\psi(r), 
\end{equation}
 which tends to $P(r)=\chi_b D_{ion}(r)$ for $r \gg \lambda_e$. $\lambda_e$ is equal to 0.4 nm for the set of parameters reproducing water properties.

 \subsection{Power expansion or Born expansion of the mean-field polarisation}
 We first recall that in general for an equation of the form
 \begin{equation}
 \tag{C1}
 \mathcal{L}[P](x) = f(x), \label{eqC1}
 \end{equation}with $\mathcal{L}[\cdot ]$ a linear operator, one can express the solution to be :
 \begin{equation}
 \tag{C2}
 P(x) = \int \mathcal{G}(x, x')f(x') \: dx', \label{eqC2}
 \end{equation}where $\mathcal{G}(x, x')$ is the Green function associated to $\mathcal{L}[\cdot ]$ satisfying
 \begin{equation}
 \tag{C3}
  \mathcal{L}[\mathcal{G}](x,x') = \delta(x-x'). \label{eqC3}
 \end{equation}The general one-dimensional mean field equation in Eq.\eqref{1dmeanfield} of the main text can be recast in the form:
 \begin{equation}
\tag{C4}
2\gamma P^3(x) + \mathcal{L}[P](x) = D(x), \label{eqC4} 
\end{equation}with $\mathcal{L}[P](x) \equiv \int dy \: M_{\parallel \: x , y}^{-1}P(y)$ for which we know the Green function to be $\chi_{_{\parallel \: lin}}(x-x')$ ({\it cf.} Eq. \eqref{gausspol}). Moving the nonlinear term on the right hand side gives rise to a recursive equation which leads to a natural expansion of the solution sometimes referred to as the Born expansion \cite{Das06}:
\begin{align}
P(x) & = \int dx_1 \: \chi_{_{\parallel \: lin}}(x-x_1) (D(x_1) - 2\gamma P^3(x_1)) \nonumber \\
& = \int dx_1 \: \chi_{_{\parallel \: lin}}(x-x_1) \Bigg[ D(x_1) + \Bigg.  \nonumber \\
 & \:\: - 2\gamma  \Bigg( \int dx_2 \: \chi_{_{\parallel \: lin}}(x_1-x_2) (D(x_2) + 2\gamma P^3(x_2))\Bigg)^3 \Bigg] \nonumber \\
& \overset{\mathcal{O}(D^3)}{=}  \int dx_1 \: \chi_{_{\parallel \: lin}}(x-x_1)D(x_1) - \int dx_1 \: \chi_{_{\parallel \: lin}}(x-x_1) \nonumber \\
& \:\:\:\: \qquad \times 2\gamma \Bigg( \int dx_2 \: \chi_{_{\parallel \: lin}}(x_1-x_2) D(x_2) \Bigg)^3. \tag{C5} \label{eqC5}
\end{align}Using the result from Eq. \eqref{basicidea} it comes:
\begin{align}
P(x) & = \chi_{_{\parallel \: \rm lin}}(q) D_0 \cos(q x) \nonumber \\
& \:\:\:\: - 2\gamma \int dx_1 \chi_{_{\parallel \: \rm lin}}(x-x_1) \chi_{_{\parallel \: \rm lin}}(q)^3 D_0^3 \cos^3(q x) \nonumber \\
& = D_0 \chi_{_{\parallel \: lin}}(q) \cos(qx) \nonumber \\
& \:\:\:\: - 2\gamma D_0^3\left(3\chi_{_{\parallel \: lin}}(q)^4\cos(q x)+\chi_{_{\parallel \: lin}}(q)^3\chi_{_{\parallel \: lin}}(3q) \right). \tag{C6} \label{eqC6}
\end{align}
\bibliographystyle{unsrt}
 \bibliography{bibnonlocnonlin6}
\end{document}